\documentclass[twocolumn]{aastex701}
\received{ }
\revised{ }
\accepted{ }

\begin{document}

\title{Investigating Possible Stellar Coronal Mass Ejections in M-type Dwarfs Using the CARMENES DR1 Spectra}

\author[orcid=0000-0002-3534-1740]{Dongtao Cao}
\affiliation{Yunnan Observatories, Chinese Academy of Sciences, Kunming 650216, China}
\affiliation{Key Laboratory for the Structure and Evolution of Celestial Objects, Chinese Academy of Sciences, Kunming 650216, China}
\affiliation{International Centre of Supernovae, Yunnan Key Laboratory, Kunming 650216, China}
\email[show]{dtcao@ynao.ac.cn}

\author{Shenghong Gu}
\affiliation{Yunnan Observatories, Chinese Academy of Sciences, Kunming 650216, China}
\affiliation{Key Laboratory for the Structure and Evolution of Celestial Objects, Chinese Academy of Sciences, Kunming 650216, China}
\affiliation{School of Astronomy and Space Science, University of Chinese Academy of Sciences, Beijing 101408, China}
\email[show]{shenghonggu@ynao.ac.cn}

\correspondingauthor{Dongtao Cao, Shenghong Gu}
\begin{abstract}
Stellar coronal mass ejections (CMEs) have significant impacts on surrounding exoplanets and stellar evolution. For M-type dwarf stars, hosting close-in habitable zones, frequent and intense stellar CMEs may render their potentially habitable planets uninhabitable. In this study, we used the CARMENES DR1 spectra to investigate potential stellar CMEs in M-type dwarfs by analyzing asymmetric features of the H${\alpha}$ profiles. Through a comprehensive visual inspection, we identified 61 distinct asymmetric features in the H${\alpha}$ line profiles across 33 M-type dwarfs. These asymmetries were associated with increased emissions in several chromospheric activity lines, indicating likely connections to flare events. The velocities of these features remain below the escape velocities of the investigated stars, except for one case where an asymmetric feature's maximum velocity approaches the star’s escape velocity. While stellar CMEs are one potential explanation, the observed features are also consistent with other flare-related plasma motions. The estimated moving plasma masses range from $10^{15}$ to $10^{19}$~g, with corresponding kinetic energies of $10^{29}$--$10^{32}$~erg. Stars with asymmetric H${\alpha}$ profiles are generally associated with stronger surface-average magnetic field ($\langle B \rangle$) and higher normalized X-ray luminosity (log$\frac{L_{X}}{L_{bol}}$). However, we find no significant correlation between the line asymmetry rate and either $\langle B \rangle$ or log$\frac{L_{X}}{L_{bol}}$.
\end{abstract}



\section{Introduction}\label{sec1}
Stellar coronal mass ejections (CMEs) are large-scale expulsions of plasma and magnetic field into interplanetary space, drawing attention for their potential effects on nearby exoplanets. CMEs originating from host stars can cause atmospheric erosion and compositional alterations in orbiting exoplanets due to increased ionizing radiation and high-energy particles \citep{Airapetian2016, Cherenkov2017, Hazra2022, Cohen2022}. Additionally, frequent CMEs may significantly contribute to stellar mass and angular momentum loss, thereby playing crucial roles in stellar evolution \citep{Aarnio2012, Osten2015}.

M-type dwarfs, the most abundant stellar population, constitute approximately 75\% of all stars in the Milky Way \citep{Henry2006}. Their low mass and low luminosity result in closer-in habitable zones, making them advantageous for detecting potentially habitable exoplanets. However, this advantage might be offset by their extreme space weather conditions. Active M-type dwarfs can produce frequent and energetic flares; some exceed the strongest solar flare on record \citep[e.g.,][]{Gunther2020, Paudel2024, Ren2026}. Solar studies show that higher-energy flares are more likely to be accompanied by CMEs \citep{Yashiro2006}. Extending this correlation to M-type dwarfs, planets located in the habitable zones of these stars would likely experience significantly more frequent and intense CMEs, potentially making them uninhabitable \citep[e.g.][]{Drake2013, Odert2017}.

Although CMEs on the Sun have been frequently detected over the past few decades, with several ones occurring daily during active phases \citep{Webb2012}, detections of stellar CMEs remain relatively rare. Current observational limitations preclude direct imaging of stellar CMEs in the same manner as doing for the Sun, necessitating reliance on other evidences for their identification. Comprehensive reviews of stellar CME detection methodologies are detailed in \citet{Osten2017}, \citet{Leitzinger2022} and \citet{Tian2023}. At present, most stellar CME candidates are identified through the detection of asymmetries in optical spectral line profiles, particularly the Balmer lines. For instance, \citet{Houdebine1990} detected a very broad blue wing enhancement in the H${\gamma}$ line profile on dMe star AD~Leo, which exhibited a projected maximum velocity of about 5800~km~s$^{-1}$. This represents the first known stellar analog to solar CMEs and is currently recognized as the fastest event documented to date. Since that time, researchers have progressively discovered potential stellar CMEs on other active stars as well \citep[see][and references therein]{Leitzinger2022}. Recently, we identified a potential flare-associated CME candidate on the RS CVn-type star II~Peg \citep{Cao2024}. This candidate exhibited as a prominent redshifted emission signature in the high-resolution H${\alpha}$ spectral profile, with a substantial bulk velocity of approximately 429 km~s$^{-1}$, which significantly exceeds the surface escape velocity of II~Peg. Moreover, several flare-associated Doppler-shifted emission components in the chromospheric line profiles of II~Peg were detected based on a long-term high-resolution spectroscopic monitoring dataset \citep{Cao2025a}. These spectral signatures are usually interpreted as eruptive prominences, which are believed to form the cores of the three-part structure observed in solar CMEs \citep{Forbes2000}. When these erupting prominences align along the observer's line of sight, known as filaments, they usually produce distinctive absorption features in stellar spectral line profiles \citep[see][]{Namekata2022, Cao2025b}. However, as noted by \citet{Leitzinger2022MNRAS}, filaments can also appear in emission around M-type dwarfs, depending on their plasma properties. 

While the majority of possible stellar CMEs documented in the literature are identified through serendipitous observations of individual events associated with specific stars, a complementary approach involves the systematic mining of archival spectroscopic datasets. This approach facilitates statistically robust investigations into stellar CME activity. For example, \citet{Vida2019} used the Polarbase archive to search for potential stellar CMEs on M-type dwarfs. Dozens of wing asymmetries in the Balmer lines during flares were detected, the fastest ones being in the order of a few hundreds of km~s$^{-1}$. \citet{Koller2021} used spectra from the SDSS DR~14 to investigate CMEs. The study encompassed spectral types F--M and identified six potential CME candidates, characterized by enhancements in either the blue or red wings, all of which were observed on M-type dwarf stars. \citet{Lu2022} utilized the LAMOST medium-resolution spectra to search for CMEs. In their study, only a limited number of events were identified to exhibit enhancements in both the red and blue wings of the Balmer lines. \citet{Leitzinger2020} used the the ESO phase~3 and Polarbase archives to explore stellar CMEs on late-type stars with spectral types F, G, and K. In more than 3700 hours of on-source time of 425 stars, no evidences of stellar CMEs could be found.

\citet{Fuhrmeister2018} conducted a high-resolution spectral survey focusing on the wings and asymmetries in the profiles of chromospheric lines in M-type dwarfs by utilizing data collected in the CARMENES (Calar Alto high-Resolution search for M dwarfs with Exo-earths with Near-infrared and optical Echelle Spectrographs) project prior to March 2017. Their investigation revealed several dozen instances of blue or red wing enhancements in the H${\alpha}$ line profiles, as well as other spectral lines. Considering the significance of M-type dwarfs to potentially habitable planets, here we present a new detailed analysis using the expanded CARMENES spectral dataset to investigate stellar CMEs on M-type dwarfs. We provide details on the data sample analyzed in Section~\ref{sec2}. Section~\ref{sec3} describes the analytical methods employed and presents the results obtained. Section~\ref{sec4} offers a comprehensive discussion of our findings. Finally, Section~\ref{sec5} summarizes the present study and outlines future works.

\begin{deluxetable*}{lccccccccccc}
\tablenum{1}
\tablecaption{Basic Parameters of the Investigated Stars Showing Asymmetric Features in the H${\alpha}$ Line Profiles and the Utilized Spectral Information in Our Analysis}
\label{tab1}
\tablewidth{0pt}
\tablehead{
\colhead{Karmn}     & \colhead{Name} & \colhead{SpT} & \colhead{$T_{eff}$} &\colhead{Mass}         & \colhead{Radius}       & \colhead{$vsini$} & \colhead{$P_{rot}$} & \colhead{$\langle$B$\rangle$} &\colhead{log$\frac{L_{X}}{L_{bol}}$} &\colhead{$v_{esc}$} & \colhead{Number of}\\
\colhead{}          & \colhead{} & \colhead{}    & \colhead{[K]}          &\colhead{[M$_{\sun}$]} & \colhead{[R$_{\sun}$]} & \colhead{[km~s$^{-1}$]}& \colhead{[day]}     & \colhead{[G]}&\colhead{}&\colhead{[km~s$^{-1}$]} & \colhead{spectra}
}
\startdata
J01019+541&  G 218-20                & M5.0 & 3070 &  0.14  &  0.16   & 30.6 & 0.2779$\pm$0.0006& \nodata & \nodata & 589 & 21\\       
J01033+623&  V388 Cas                & M5.0 & 3060 &  0.23  &  0.24   & 10.5 & 1.02$\pm$0.01 & 4800$\pm$300 & -2.97$\pm$0.05 & 617 & 27 \\ 
J01125-169&  YZ Cet                  & M4.5 & 3230 &  0.14  &  0.15   &  \textless~2 &69.2$\pm$2.4& \nodata & -3.46$\pm$0.044 & 609 & 82\\ 
J02088+494&  G 173-39                & M3.5 & 3260 &  0.37  &  0.42   & 24.1 &0.74759$\pm$0.00019& 4900$\pm$1000 & -2.85$\pm$0.072 & 591 & 17\\ 
J02519+224&  RBS 365                 & M4.0 & 2990 &  0.25  &  0.61   & 27.2 &0.85757$\pm$0.00003& \nodata & -2.21$\pm$0.023 & 403 & 15\\ 
J04198+425&  LSR J0419+4233          & M8.5 & 2400 &  0.09  &  0.11   &  3.6 &0.99& \nodata & \nodata & 570 & 46\\ 
J05062+046&  RX J0506.2+0439         & M4.0 & 3010 &  0.26  &  0.62   & 29 &0.8650$\pm$0.0006& 3070$\pm$490 & -3.12$\pm$0.063 & 408 & 13\\ 
J05084-210&  2MASS J05082729-2101444 & M5.0 & 3233 &  0.15  &  0.68   & 25.2 &0.28& \nodata & -3.17$\pm$0.108 & 630 & 26\\ 
J05337+019&  V371 Ori                & M3.0 & 3430 &  0.49  &  0.49   &  9.8 &0.60& \nodata & -3.01$\pm$0.043 & 630 & 14\\ 
J06318+414&  LP 205-44               & M5.0 & 3260 &  0.35  &  0.36   & 54 &0.29952$\pm$0.00007& 2200$\pm$1150 & -2.96$\pm$0.096 & 621 &36\\ 
J06574+740&  2MASS J06572616+7405265 & M4.0 & 3137 &  0.27  &  0.28   & 32 &0.61& 3330$\pm$830  & -2.93$\pm$0.053 & 619 & 12\\ 
J07319+362N& BL Lyn                  & M3.5 & 3384 &  0.41  &  0.40   & 1.7  &16.4$\pm$0.3& 2530$\pm$140  & -2.75$\pm$0.021 & 638 & 38\\
J07472+503&  2MASS J07471385+5020386 & M4.0 & 3250 &  0.27  &  0.28   & 10.3 &1.32$\pm$0.01& 2940$\pm$290  & -3.76$\pm$0.122 & 619 & 16\\ 
J08298+267&  DX Cnc                  & M6.5 & 3000 &  0.10  &  0.12   & 11.5 &0.45900$\pm$0.00001& 2680$\pm$440  & \nodata & 575 & 35\\
J08413+594&  LP 90-18                & M5.5 & 3140 &  0.12  &  0.14   & 2.0  &83.27& 1280$\pm$210  & \nodata & 583 & 186\\
J10196+198&  AD Leo                  & M3.0 & 3455 &  0.45  &  0.43   & 3.3  &2.24 & 3570$\pm$90   & -3.08$\pm$0.023 & 644 & 46\\
J10564+070&  CN Leo                  & M6.0 & 3071 &  0.10  &  0.11   & 0.0  &2.704$\pm$0.003& 3010$\pm$160  & -3.54$\pm$0.049 & 601 & 77\\
J10584-107&  LP 731-76               & M5.0 & 3218 &  0.19  &  0.21   & 2.8  &27.87& 4100$\pm$170  & -3.30$\pm$0.015 & 599 & 53\\
J11055+435&  WX UMa                  & M5.5 & 3280 &  0.10  &  0.13   & 3.5  & 0.78$\pm$0.02 & 6880$\pm$140  & -3.15$\pm$0.063 & 553 & 37\\ 
J11474+667&  1RXS J114728.8+664405   & M5.0 & 3171 &  0.29  &  0.30   & 3.1  & 13.3 & 4770$\pm$140  & -3.67 & 619 & 43\\ 
J13536+776&  RX J1353.6+7737         & M4.0 & 3200 &  0.30  &  0.30   & 9.7  &1.23$\pm$0.01& 2630$\pm$400  & -3.13$\pm$0.005 & 630 & 25\\ 
J13591-198&  LP 799-7                & M4.0 & 3261 &  0.26  &  0.27   & 4.2  & 3.32 & 1910$\pm$240  & -3.21$\pm$0.061 & 618 & 17\\
J15218+209&  OT Ser                  & M1.5 & 3700 &  0.52  &  0.52   & 4.1  &3.37$\pm$0.01& 3360$\pm$100  & -3.20$\pm$0.035 & 630 & 54\\
J15305+094&  NLTT 40406              & M5.5 & 3073 &  0.11  &  0.13   & 13.4 &0.3048$\pm$0.0006& 4780$\pm$600  & \nodata & 580 & 14\\
J16570-043&  LP 686-27               & M3.5 & 3206 &  0.25  &  0.26   & 9.7  &0.547$\pm$0.001& 3660$\pm$380  & -3.10$\pm$0.05 & 618 & 15\\ 
J18022+642&  LP 71-82                & M5.0 & 3210 &  0.16  &  0.18   & 12.9 &0.28027$\pm$0.00002& 4930$\pm$310  & -3.18$\pm$0.013 & 594 & 29\\
J18482+076&  G 141-36                & M5.0 & 3180 &  0.14  &  0.16   & 3.6  &2.76$\pm$0.01& 1190$\pm$230  & -3.65$\pm$0.121 & 589 & 50\\
J20451-313&  AU Mic                  & M0.5 & 3768 &  0.61  &  0.86   &  8.2 & 4.89 & 3010$\pm$220  & -2.85$\pm$0.009 & 531 & 100\\ 
J22114+409&  1RXS J221124.3+410000   & M5.5 & 3130 &  0.16  &  0.18   &  2.0 &30.0$\pm$1.3& 1880$\pm$210  & -2.94$\pm$0.083 & 594 & 58\\ 
J22231-176&  LP 820-12               & M4.5 & 3228 &  0.18  &  0.19   &  2.3 & 4.57 & 2150$\pm$240  & -3.23$\pm$0.072 & 613 & 10\\ 
J22468+443&  EV Lac                  & M3.5 & 3310 &  0.34  &  0.34   &  4.1 &4.38$\pm$0.03& 4320$\pm$110  & -2.59$\pm$0.008 & 630 & 106 \\ 
\enddata 
\end{deluxetable*}
\begin{deluxetable*}{lccccccccccc}
\tablenum{1}
\tablecaption{Continued}
\label{tab1}
\tablewidth{0pt}
\tablehead{
\colhead{Karmn}     & \colhead{Name} & \colhead{SpT} & \colhead{$T_{eff}$} &\colhead{Mass}         & \colhead{Radius}       & \colhead{$vsini$} & \colhead{$P_{rot}$} & \colhead{$\langle$B$\rangle$} &\colhead{log$\frac{L_{X}}{L_{bol}}$} &\colhead{$v_{esc}$} & \colhead{Number of}\\
\colhead{}          & \colhead{} & \colhead{}    & \colhead{[K]}          &\colhead{[M$_{\sun}$]} & \colhead{[R$_{\sun}$]} & \colhead{[km~s$^{-1}$]}& \colhead{[day]}     & \colhead{[G]}&\colhead{}&\colhead{[km~s$^{-1}$]} & \colhead{spectra}
}
\startdata
J22518+317&  GT Peg                  & M3.0 & 3170 &  0.35  &  0.54   & 12.7 &1.63$\pm$0.01& 3870$\pm$340  & -3.03$\pm$0.029 & 507 & 12  \\ 
J23548+385&  RX J2354.8+3831         & M4.0 & 3263 &  0.31  &  0.32   &  3.6 &4.70$\pm$0.04& 4900$\pm$130  & -2.86$\pm$0.053 & 620 & 12\\ 
\enddata 
\tablecomments{The spectral type (Spt), effective temperature ($T_{eff}$), mass, radius, rotational velocity ($vsini$), rotation period ($P_{rot}$), surface-average magnetic field ($\langle B \rangle$), and normalized X-ray luminosity (log$\frac{L_{X}}{L_{bol}}$) are derived from the studies conducted by \citet{Reiners2018}, \citet{Shulyak2019}, \citet{Reiners2022}, \citet{Ribas2023}, and \citet{Shan2024}. It should be noted that certain measured values for $\langle B \rangle$ and log$\frac{L_{X}}{L_{bol}}$ have not been obtained. The stellar surface escape velocity ($v_{esc}$) is calculated using the formula $v_{esc}~=~630(\frac{M_{\star}}{M_{\sun}})^{1/2}(\frac{R_{\star}}{R_{\sun}})^{-1/2}$~km~s$^{-1}$. The ``Number of spectra'' column indicates the total count of analyzed spectra for each star.}
\end{deluxetable*}
\begin{figure}
\centering
\includegraphics[width=8.75cm,height=10cm]{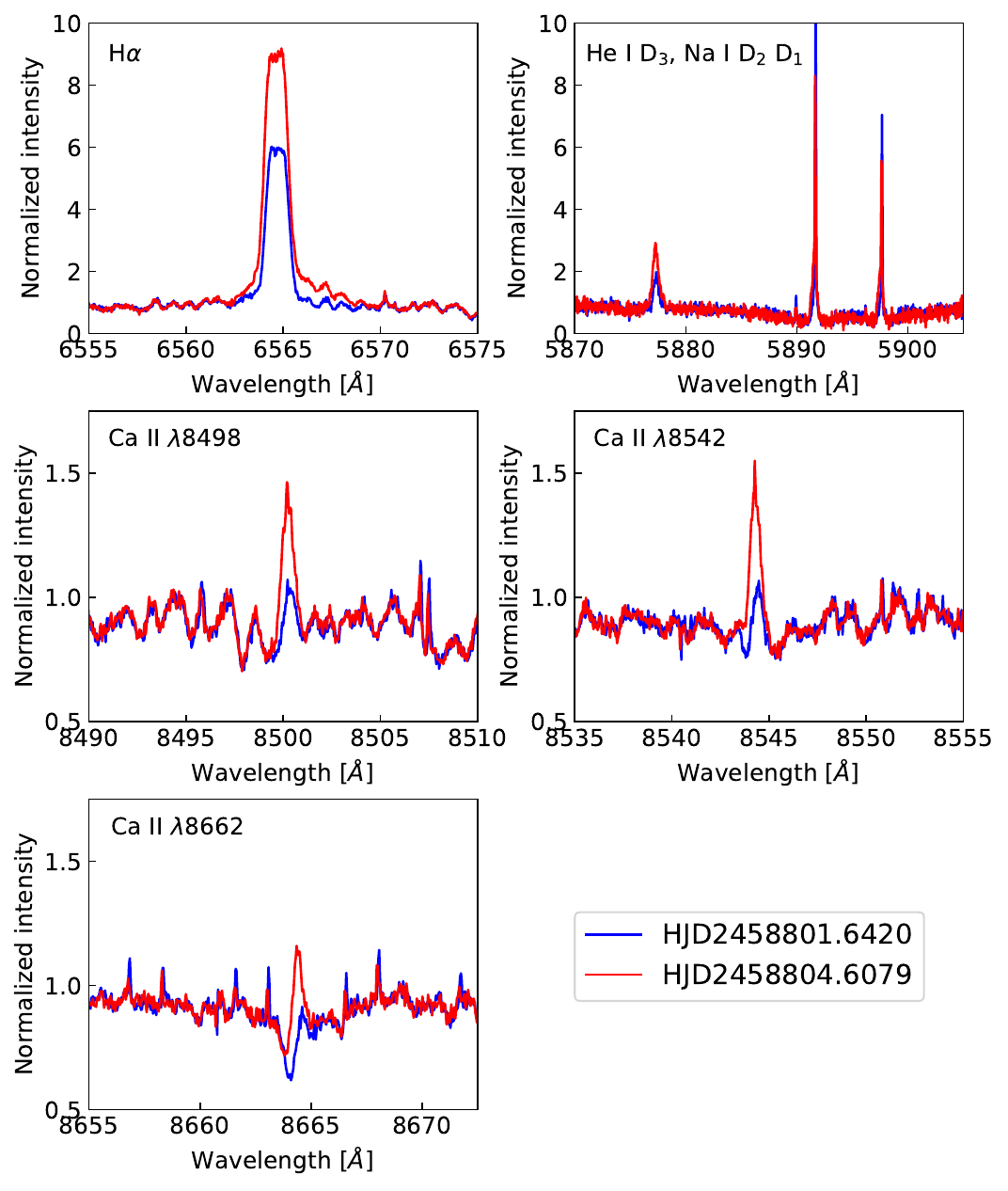}
\caption{Examples of the spectral line profiles probably associated with a flare (indicated by the red spectrum observed on HJD2458804.6079) in the H${\alpha}$, $\mbox{He~{\sc i}}$ D$_{3}$, and $\mbox{Na~{\sc i}}$ D$_{1}$ and D$_{2}$ doublet, as well as $\mbox{Ca~{\sc ii}}$ lines at $\lambda$8498, $\lambda$8542, and $\lambda$8662 for the investigated star V388~Cas (Karmn~J01033+623). The flaring spectrum exhibits a pronounced asymmetric feature in the red wing of the H${\alpha}$ line profile and demonstrates stronger emission than the spectrum observed from the closest night (the blue spectrum observed on HJD2458801.6420).}
\label{Fig1}
\end{figure}

\section{Data sample}\label{sec2}
All spectra analyzed in this paper were obtained from the CARMENES Guaranteed Time Observations (GTO) data release~1 (DR1) \citep{Ribas2023}, which spans the period from 2016 to 2020. During this period, a total of 19633 spectra were collected for a sample of 362 targets. The primary scientific aim of the CARMENES project is to conduct a search for low-mass planets that orbit M-type dwarfs in their habitable zones using the 3.5m telescope at the Calar Alto Observatory in Almer\'{i}a, Spain \citep{Quirrenbach2014}. The CARMENES spectrograph features a two-channel, fiber-fed design that spans a wavelength range from 5200~$\AA$ to 9600~$\AA$ with a spectral resolution of R~$\approx$~94,600 in the visual channel (VIS) and from 9600~$\AA$ to 17100~$\AA$ with R~$\approx$~80,400 in the near-infrared channel (NIR) \citep{Quirrenbach2016}. In the CARMENES DR1, only the spectra in the VIS channel are publicly available online. 

The exposure times for the CARMENES spectra generally ranged from under 100 seconds to several hundred seconds, with a maximum value of 1800 seconds. This variation primarily depended on the brightness of the stars. Due to considerations of observability and scientific reasons, the observing cadence is random and non-uniform \citep{Ribas2023}. The CARMENES spectra were reduced by the standard reduction pipeline called CARACAL \citep{Zechmeister2014, Caballero2016}, which corrects for bias, flat-field, and cosmic rays, and extracts one-dimensional, wavelength-calibrated spectra. For our analysis, the spectra are corrected for barycentric and stellar radial velocity shifts, following the methodology described by \citet{Fuhrmeister2018}. For an investigated star LSR~J0419+4233 (Karmn~J04198+425), the absolute radial velocity was not provided, likely due to its low signal-to-noise ratio. Therefore, we correct its radial velocity by fitting the H${\alpha}$ emission line. Finally, for continuum normalization of the analyzed spectra, we have employed a low-order polynomial fit to the observed continuum utilizing the CONTINUUM task of the IRAF\footnote{IRAF is distributed by the National Optical Astronomy Observatories, which is operated by the Association of Universities for Research in Astronomy (AURA), Inc., under cooperative agreement with the National Science Foundation.} package.

In the CARMENES sample, a considerable proportion exhibits either inactivity or low-activity levels. Taking into the consideration of the strong correlation between CMEs and flaring activity, as demonstrated by solar observations, we have selected over 90 stars exhibiting pronounced activity for targeted investigation. These stars are identified as potential sources of frequent flares and consequently CMEs, which are chosen based on distinct characteristics observed in their H${\alpha}$ spectral lines through a visual inspection. This includes those that consistently exhibit prominent emission features above the continuum and those that typically present filled H${\alpha}$ emission but occasionally display clear emission features likely attributable to flare events. Furthermore, certain binary stars, such as 2MASS~J07000682-1901235 (Karmn~J07001-190), are excluded from our analysis despite exhibiting asymmetric H${\alpha}$ profiles. This exclusion is necessitated by the potential that the asymmetric features may originate from the emission lines of the secondary components in binary systems, while also ensuring the sample remains homogeneous, aligning with the strategy adopted by \citet{Vida2019}.

\section{Analysis and results}\label{sec3}
For each target star, we have generated spectral plots of the H${\alpha}$ region and conducted a visual inspection to identify asymmetric profiles. Through this analysis, we have detected 61 distinct asymmetric features in the H${\alpha}$ line profiles of 33 targets (basic parameters are summarized in Table~\ref{tab1}). Because the CARMENES observations lack continuous time series, most asymmetric features were captured in single-epoch spectra, making it challenging to definitively associate these features with flare events solely based on snapshot observations. However, it is important to note that these asymmetric features were primarily linked to enhanced emissions in chromospheric activity lines when compared to the spectra obtained from the closest nights. This suggests they are likely associated with flare activities. These enhancements are clearly reflected in the H${\alpha}$ line profiles, as well as in the \mbox{He~{\sc i}}~D$_{3}$, \mbox{Ca~{\sc ii}}~$\lambda$8498, \mbox{Ca~{\sc ii}}~$\lambda$8542, and \mbox{Ca~{\sc ii}}~$\lambda$8662 lines (as examples shown in Figure~\ref{Fig1}). In some spectra with asymmetric H${\alpha}$ profiles, the observed spectral enhancements were relatively weak. This could be attributed to observations taken during the decay phase of flares, though we can not exclude the possibility that rotational modulation of chromospheric active regions contributed to the observed variability.

To accurately determine the velocities and intensities of asymmetric features in the H${\alpha}$ line profile, it is essential to subtract a quiescent reference spectrum from the flaring spectrum.  Previous studies, such as those conducted by \citet{Vida2019} and \citet{Fuhrmeister2018}, have employed different methodologies for defining the quiescent states of stars. Specifically, \citet{Vida2019} utilized the average spectrum of each star derived by excluding clearly active spectra, whereas \citet{Fuhrmeister2018} selected the spectrum exhibiting the lowest activity for each star. In light of the extensive temporal coverage offered by the CARMENES dataset, and taking into account the intrinsic chromospheric variability characteristics of active M-type dwarfs, we have utilized the temporally closest spectra as quiescent references for our analysis of flaring spectra. These spectra closest in time are more likely to accurately represent the stars' normal states close to the flare events, thereby minimizing the influences of long-term stellar variability. However, it is important to note that the influence of stellar rotation modulation cannot be entirely eliminated. These quiescent reference spectra predominantly source from observational periods prior to flare events; in instances where suitable pre-flare observations are not available, we resort to use post-flare quiescent spectra. All quiescent reference spectra are characterized by lack of stronger emission features and do not display significant enhancements or absorption characteristics in their wing regions, as illustrated in Figure~\ref{Fig1}. Subsequently, we subtract the quiescent reference spectrum from the flaring spectrum to derive a residual profile for each case and then make further analysis.

\begin{figure*}
\centering
\includegraphics[width=11.cm,height=5.25cm]{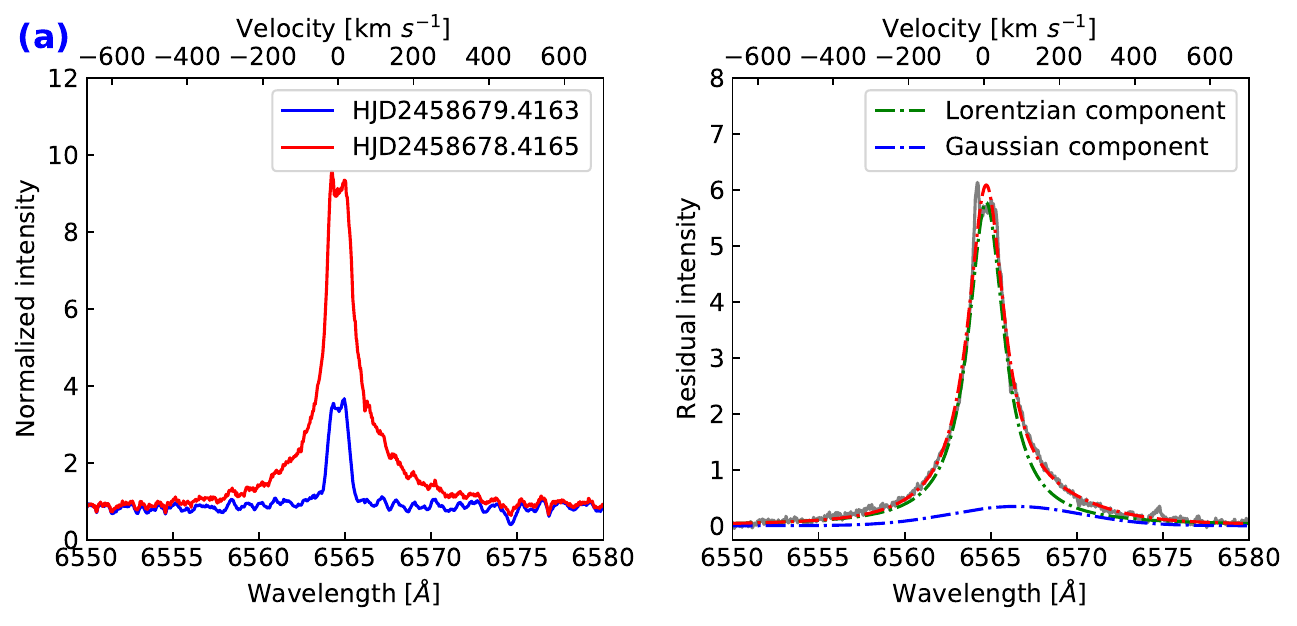}
\includegraphics[width=11.cm,height=5.25cm]{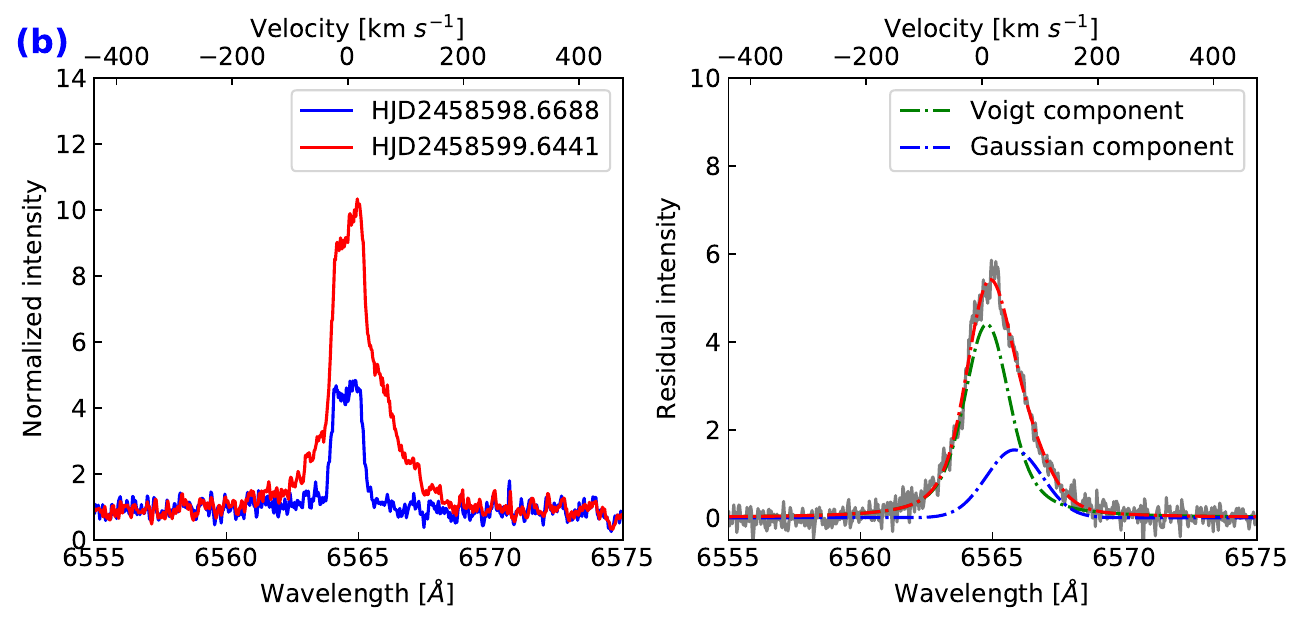}
\includegraphics[width=11.cm,height=5.25cm]{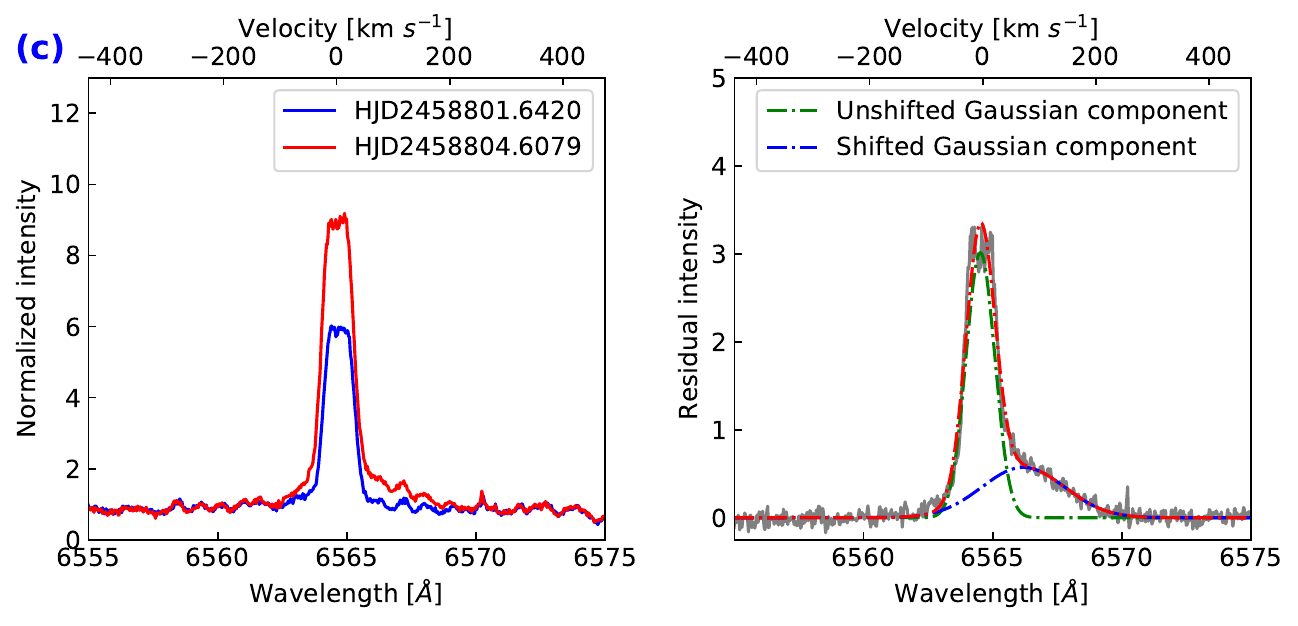}
\includegraphics[width=11.cm,height=5.25cm]{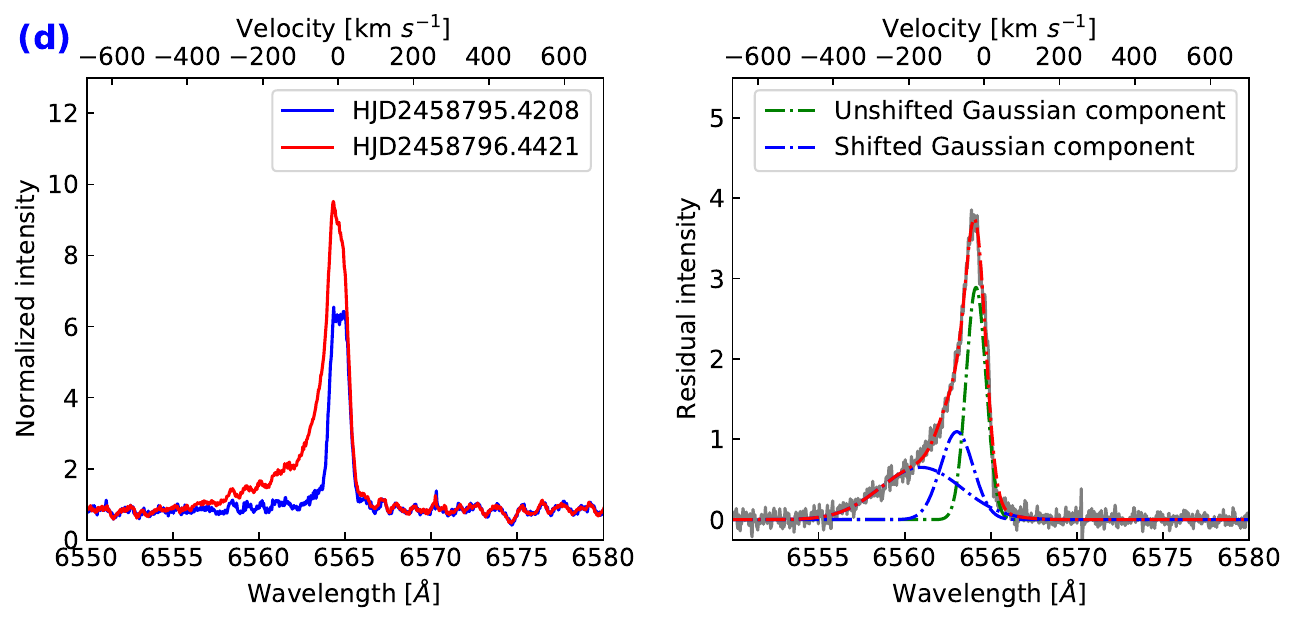}
\caption{Examples of the asymmetric profiles. Left panels: The H${\alpha}$ profiles displaying asymmetric features (red line) are compared with the profiles in a quiescent state (blue line) for the following stars: (a) RX~J1353.6+7737 (Karmn~J13536+776), (b) 1RXS~J221124.3+410000 (Karmn~J22114+409), and (c)--(d) V388~Cas (Karmn~J01033+623). Right panels: Residual profiles (gray line) and their fittings are presented. (a) A Lorentzian profile combined with a Gaussian profile, (b) a Voigt profile combined with a Gaussian profile, (c) a Gaussian profile combined with another Gaussian profile, (d) a Gaussian profile combined with two additional Gaussian profiles. The red dash-dotted lines correspond to the sum of these profiles.}
\label{Fig2}
\end{figure*}

Generally, flaring H${\alpha}$ spectra display broader profiles that resemble Lorentzian-like shapes, attributed to the significant effects of Stark broadening \citep{Kowalski2017, Namekata2020}. \citet{Wu2022} utilized a Voigt function that includes a Lorentzian component to model the flaring H${\alpha}$ profile of a M-type dwarf. Similar approaches had also been employed by \citet{Namekata2022} and \citet{Namizaki2023}. In our analysis, some residual H${\alpha}$ profiles display distinct Lorentzian-like characteristics, even though they might be fitted using a narrow Gaussian emission component alongside a broad Gaussian emission component, as adopted by \citet{Fuhrmeister2018}. However, in our analysis, we firstly prefer to employ either a Lorentzian or Voigt component to model the symmetric features of these residual profiles while incorporating a Gaussian component to account for any potential asymmetric features (see Figure~\ref{Fig2}a and \ref{Fig2}b ). In instances where the symmetric features of the residual H${\alpha}$ profiles do not demonstrate the anticipated Lorentzian-like behavior and are not better fitted by employing Lorentzian or Voigt components, we have utilized Gaussian components for modeling. Additionally, other Gaussian components are employed to address any possible asymmetric features (see an example in Figure~\ref{Fig2}c). Moreover, it is notably that in specific cases where a single Gaussian component proves inadequate for fitting the asymmetric feature accurately, two Gaussian components are necessary for an appropriate fit, as shown in Figure~\ref{Fig2}d. These cases might indicate the complex physical processes occurring during the flares.

In this study, we have focused exclusively on spectra displayed pronounced asymmetric H${\alpha}$ lines. Spectra exhibiting enhanced H${\alpha}$ emission and occasionally more broadening profiles, but lacking prominent asymmetry, are excluded from our current analysis; some extreme cases may be considered for investigation in future research. In their analysis using a subset of the data adopted in the present study, \citet{Fuhrmeister2018} identified 63 asymmetric features in the H$\alpha$ line profiles—a number slightly higher than that found here. The difference mainly arises from the distinct methodologies employed. A considerable fraction of the asymmetric features reported by them correspond to relatively low velocity shifts; for example, 22\% of those features show shifts smaller than $\pm$10~km~s$^{-1}$. In cases where both the narrow and broad Gaussian components are fitted with minor shifts, a single Lorentzian or Voigt profile can adequately describe the line shape. In our analysis, such profiles are treated as symmetric.

Table~\ref{tab2} presents the observing information and several key measurements for each detected asymmetric feature. This includes the heliocentric Julian date (HJD) corresponding to the spectrum that exhibits an asymmetric H${\alpha}$ profile, the HJD of the quiescent reference spectrum, and the exposure time (Exp.time) for the spectrum exhibiting this asymmetric H${\alpha}$ profile. Additionally, it encompasses measurements of equivalent width (EW), bulk velocity (V$_{bulk}$), and maximum velocity (V$_{max}$) associated with the asymmetric feature. The bulk velocity represents the centroid velocity derived from Gaussian fitting of the asymmetric feature, while the maximum velocity is calculated as V$_{max}$~=~V$_{bulk}$~$\pm$~2$\sigma$, where $\sigma$ is the Gaussian width parameter obtained from the same fit (i.e., the standard deviation of the fitted Gaussian profile). For blueshifted asymmetries, this value is assigned a negative sign (``-''), whereas for redshifted asymmetries, it is assigned a positive sign (``+''). Additionally, we have indicated in the table which function is employed to fit the residual H${\alpha}$ profile. In cases where two Gaussian components are required to adequately model an asymmetric feature, we have identified them as one event and provided measurements for each component separately.

Based on the EWs of the asymmetric features observed in the H${\alpha}$ line profiles, we can estimate the masses of the moving plasma responsible for these asymmetries. By following \citet{Houdebine1990} and \citet{Koller2021}, we calculate the lower limits on the masses following
\begin{eqnarray}
M_{CME}~\geqslant~\frac{4{\pi}d^{2}f_{line}m_{H}}{A_{ji}h\nu_{ji}P_{esc}}\frac{N_{tot}}{N_{j}},
\end{eqnarray}
in which $d$ represents the distance from the star, $f_{line}$ denotes the corresponding line flux associated with the asymmetric feature, and $m_{H}$ is the mass of a hydrogen atom. The term $N_{tot}/N_{j}$ indicates the ratio between the number density of hydrogen atoms and that at excited level $j$. Additionally, $h$ refers to Planck's constant, while $\nu_{ji}$ and $A_{ji}$ represent the frequency and Einstein coefficient for spontaneous decay from level $j$ to level $i$, respectively. Lastly, $P_{esc}$ signifies the escape probability. Detailed parameter settings can be found in \citet{Koller2021}. Utilizing the method described by \citet{Wu2022}, we have calculated the luminosity of these asymmetric features to replace \(4{\pi}d^{2}f_{\text{{line}}}\), subsequently deriving their masses as detailed in Table~\ref{tab2}. Furthermore, by employing these derived masses alongside bulk velocities of the asymmetric features, we have computed their kinetic energies for moving plasma; these results are also presented in Table~\ref{tab2}.

\begin{deluxetable*}{lccccccccccc}
\tablenum{2}
\tablecaption{Measurements of the Detected H${\alpha}$ Line Asymmetries\label{tab2}}
\tablewidth{150pt}
\tablehead{
\colhead{Karmn} & \colhead{No} & \colhead{HJD} & \colhead{HJD} & \colhead {Exp.time} & \colhead{Enhanced} & \colhead{Fitted} & \colhead{$EW_{asy}$} & \colhead{$V_{\rm bulk}$}  & \colhead{$V_{\rm max}$}        &\colhead{Mass}  & \colhead{$E_{\rm kin}$}\\
\colhead{}     & \colhead{}   & \colhead{}  &\colhead{of Reference} & \colhead{[s]} & \colhead{Wing}  & \colhead{Function}  &  [\AA] & \colhead{[km~s$^{-1}$]}  &  \colhead{[km~s$^{-1}$]} &\colhead{[g]}   & \colhead{[erg]}
}
\startdata
J01019+541  &  1  & 2457761.3231   &2457749.3893  & 1802 &Red      & G + G & 3.210  &  $23.4 \pm 2.2$  &  257  &  $2.3\times10^{17}$ & $6.4\times10^{29}$\\
            &  2  & 2457945.6608   &2457956.5731  & 1801 &Blue     & G + G &0.349  & $-81.8 \pm 1.0$  & -114  & $2.6\times10^{16}$ & $8.5\times10^{29}$\\
J01033+623  &  1  &  2457749.4135  &2457755.4148  & 1589 &Red      & G + G & 0.277  &  $96.4 \pm 3.4$  &  163  & $4.5\times10^{16}$ & $2.1\times10^{30}$\\
            &  2  &  2457814.3242  &2457822.3202  & 1802 &Far-red  & G + 2*G & 0.534  & $245.8 \pm 11.8$ &  368  & $8.6\times10^{16}$ & $2.6\times10^{31}$\\
            &     &                &              & &Red      &         & 2.417  &  $57.8 \pm 6.0$  &  220  & $3.9\times10^{17}$ & $6.5\times10^{30}$\\
            &  3  &  2458796.4421  &2458795.4208  & 1801 &Far-blue & G + 2*G & 3.870  &$-168.0 \pm 6.3$  & -385  & $6.2\times10^{17}$ & $8.8\times10^{31}$\\
            &     &                &              & &Blue     &         & 2.589  & $-72.3 \pm 10.8$ & -159  & $4.2\times10^{17}$ & $1.1\times10^{31}$\\
            &  4  &  2458804.6079  &2458801.6420  & 1802&Red      & G + G & 2.299  &  $71.5 \pm 3.9$  &  218  & $3.7\times10^{17}$ & $9.4\times10^{30}$\\
            &  5  &  2458832.3774  &2458829.5145  & 1801&Red      & G + G & 2.836  &  $90.3 \pm 1.8$  &  244  & $4.6\times10^{17}$ & $1.9\times10^{31}$\\
J01125-169  &  1  &  2457961.6737  &2457960.6860  & 1268&Blue     & V + G & 1.121  & $-25.3 \pm 2.6$  & -182  & $1.0\times10^{17}$ & $3.3\times10^{29}$\\
            &  2  &  2458487.2867  &2458486.3116  & 1576&Red      & G + G & 0.302  & $101.0 \pm 3.6$  &  161  & $2.8\times10^{16}$ & $1.4\times10^{30}$\\
J02088+494  &  1  &  2457691.5388  &2457688.5124  & 1060&Red      & G + G & 0.172  & $100.7 \pm 7.3$  &  164  & $1.3\times10^{17}$ & $6.7\times10^{30}$\\
            &  2  &  2457987.5996  &2457985.5565  & 1801&Red      & L + G & 1.368  &  $20.2 \pm 2.0$  &  150  & $1.0\times10^{18}$ & $2.1\times10^{30}$\\
J02519+224  &  1  &  2457693.5223  &2457690.5988  & 1326&Far-blue & G + 2*G & 2.961  &$-149.4 \pm 1.8$  & -404  & $2.6\times10^{18}$ & $2.9\times10^{32}$\\
            &     &                &              & &Blue     &         & 0.410  & $-77.7 \pm 1.7$  & -158  & $3.6\times10^{17}$ & $1.1\times10^{31}$\\
J04198+425  &  1  &  2458697.6303  &2458745.6896  & 1801&Blue     & G + G &106.101 & $-13.2 \pm 1.5$  & -197  & $5.0\times10^{17}$ & $4.4\times10^{29}$\\
            &  2  &  2458849.4605  &2458845.3959  & 1801&Red      & G + G & 2.557  &  $84.6 \pm 4.4$  &  138  & $1.2\times10^{16}$ & $4.3\times10^{29}$\\
J05062+046  &  1  &  2458135.4122  &2458124.4001  & 1801&Blue     & G + G & 0.847  & $-89.1 \pm 0.9$  & -160  & $8.1\times10^{17}$ & $3.2\times10^{31}$\\
J05084-210  &  1  &  2458094.5251  &2458092.5389  & 1801&Blue     & G + G & 16.216 & $-93.1 \pm 0.8$  & -245  & $1.5\times10^{18}$ & $6.5\times10^{31}$\\
J05337+019  &  1  &  2458160.3592  &2458167.3267  & 1801&Red      & G + G & 0.644  &  $51.3 \pm 14.1$ &  146  & $9.4\times10^{17}$ & $1.2\times10^{31}$\\
J06318+414  &  1  &  2457823.4526  &2457832.3790  & 1802&Red      & V + G & 8.648  & $195.7 \pm 0.4$  &  363  & $3.1\times10^{18}$ & $6.0\times10^{32}$\\
            &  2  &  2458057.6512  &2458055.6349  & 1802&Blue     & V + G & 7.845  &$-122.5 \pm 0.3$  & -210  & $2.8\times10^{18}$ & $2.1\times10^{32}$\\
J06574+740  &  1  &  2458023.5941  &2458035.6522  & 1802&Red      & V + G & 0.483  & $100.2 \pm 4.3$  & 210  & $1.3\times10^{17}$ & $6.3\times10^{30}$\\
J07319+362N &  1  &  2457449.3949  &2457444.4472  & 1201&Red      & V + G & 0.518  & $113.0 \pm 9.2$  &  254  & $4.6\times10^{17}$ & $2.9\times10^{31}$\\
J07472+503  &  1  &  2458081.5758  &2458109.5598  & 1802&Blue     & V + G & 0.620  & $-83.6 \pm 9.6$  & -195  & $2.1\times10^{17}$ & $7.2\times10^{30}$\\
            &  2  &  2458882.4574  &2458891.4284  & 1801&Blue     & V + G & 0.837  & $-51.5 \pm 4.2$  & -245  & $2.8\times10^{17}$ & $3.7\times10^{30}$\\
            &  3  &  2458897.3860  &2458894.4499  & 1801&Red      & V + G & 0.295  &  $86.6 \pm 23.4$ & 383   & $9.8\times10^{16}$ & $3.7\times10^{30}$\\ 
            &  4  &  2458903.4274  &2458894.4499  & 1801&Red      & V + G & 0.282  &  $90.8 \pm 13.9$ &  208  & $9.4\times10^{16}$ & $3.9\times10^{30}$\\
J08298+267  &  1  &  2459177.6384  &2459183.6036  & 1802&Red      & G + G & 7.569  &  $43.4 \pm 5.1$  &  235  & $2.6\times10^{17}$ & $2.5\times10^{30}$\\
            &  2  &  2458543.5011  &2458539.6104  & 1801&Red      & G + G & 1.515  &  $96.2 \pm 8.1$  &  232  & $5.3\times10^{16}$ & $2.4\times10^{30}$\\
J08413+594  &  1  &  2458142.6181  &2458141.5389  & 1801&Blue     & V + G & 0.342  & $-84.6 \pm 2.5$  &  -130 & $2.2\times10^{16}$ & $2.5\times10^{30}$\\
J10196+198  &  1  &  2458209.4747  &2458209.3553  & 167 &Red      & G + G & 0.476  &  $40.3 \pm 5.8$  &   253 & $5.6\times10^{17}$ & $4.5\times10^{30}$\\ 
J10564+070  &  1  &  2457761.3231  &2457749.3893  & 511 &Red      & G + G & 1.654  &  $46.5 \pm 21.5$ &  126  & $5.7\times10^{16}$ & $6.2\times10^{29}$\\
J10584-107  &  1  &  2458491.6338  &2458486.6952  & 1803&Blue     & V + G & 1.481  & $-78.2 \pm 6.3$  & -205  & $2.6\times10^{17}$ & $7.9\times10^{30}$\\
\enddata 
\end{deluxetable*}
\begin{deluxetable*}{lccccccccccc}
\tablenum{2} 
\tablecaption{Continued}
\tablewidth{150pt}
\tablehead{
\colhead{Karmn} & \colhead{No} & \colhead{HJD} & \colhead{HJD} & \colhead {Exp.time} & \colhead{Enhanced} & \colhead{Fitted} & \colhead{$EW_{asy}$} & \colhead{$V_{\rm bulk}$}  & \colhead{$V_{\rm max}$}        &\colhead{Mass}  & \colhead{$E_{\rm kin}$}\\
\colhead{}     & \colhead{}   & \colhead{}  &\colhead{of Reference} & \colhead{[s]} & \colhead{Wing}  & \colhead{Function}  &  [\AA] &\colhead{[km~s$^{-1}$]}  &  \colhead{[km~s$^{-1}$]} &\colhead{[g]}   & \colhead{[erg]}
}
\startdata
J11055+435  &  1  &  2458479.7479  &2458474.6781  & 1802&Red      & G + G & 5.500  &  $60.6 \pm 4.5$  &  207  & $4.2\times10^{17}$ & $7.7\times10^{30}$\\
            &  2  &  2458525.6497  &2458530.6597  & 1802&Far-red  & G + 2*G & 1.686  & $230.9 \pm 25.2$ &  436  & $1.3\times10^{17}$ & $3.4\times10^{31}$\\
            &     &                &              & &Red      &       & 2.181  & $105.2 \pm 1.5$  &  193  & $1.7\times10^{17}$ & $9.2\times10^{30}$\\
J11474+667  &  1  &  2458852.7308  &2458831.6285  & 1801&Red      & L + G & 4.510  & $109.8 \pm 20.4$ &  316  & $1.5\times10^{18}$ & $8.8\times10^{31}$\\
            &  2  &  2457762.5596  &2457759.6172  & 1801&Blue     & V + G & 10.434 & $-24.8 \pm 3.8$  & -439  & $3.4\times10^{18}$ & $1.0\times10^{31}$\\
J13536+776  &  1  &  2457558.4660  &2457529.4688  & 1800&Blue     & G + G & 0.550  & $-82.4 \pm 5.9$  & -192  & $1.9\times10^{17}$ & $6.4\times10^{30}$\\
            &  2  &  2457911.4554  &2457917.5654  & 1801&Red      & V + G & 0.183  &  $38.3 \pm 19.2$ &  109  & $6.3\times10^{16}$ & $4.6\times10^{29}$\\
            &  3  &  2458678.4165  &2458679.4163  & 1801&Red      & L + G & 3.285  &  $83.6 \pm 8.6$  &  431  & $1.1\times10^{18}$ & $3.9\times10^{31}$\\
            &  4  &  2458877.7292  &2458880.6007  & 1801&Red      & V + G & 0.089  & $116.5 \pm 10.9$ &  205  & $3.1\times10^{16}$ & $2.1\times10^{30}$\\
            &  5  &  2458895.6066  &2458893.6291  & 1800&Blue     & V + G & 0.912  & $-38.5 \pm 5.2$  & -212  & $3.1\times10^{17}$ & $2.3\times10^{30}$\\
J13591-198  &  1  &  2458904.6416  &2458895.6997  & 1801&Blue     & V + G & 0.443  & $-52.4 \pm 24.5$ & -193  & $1.4\times10^{17}$ & $1.9\times10^{30}$\\
J15218+209  &  1  &  2457752.7037  &2457755.7247  & 373 &Red      & V + G & 0.176  & $131.5 \pm 30.9$ &  293  & $6.1\times10^{17}$ & $5.3\times10^{31}$\\
            &  2  &  2457950.5052  &2457950.4156  & 867 &Red      & V + G & 0.595  &  $26.4 \pm  4.9$ &  242  & $2.1\times10^{18}$ & $7.2\times10^{30}$\\
J15305+094  &  1  &  2457954.4187  &2457956.4038  & 1801 &Blue     & G + G & 0.282  & $-23.7 \pm  0.3$ &  -38  & $1.4\times10^{16}$ & $3.9\times10^{28}$\\
            &  2  &  2458296.5191  &2458296.4562  & 1801 &Red      & G + G & 1.316  &  $37.6 \pm 4.4$  &  220  & $6.4\times10^{16}$ & $4.5\times10^{29}$\\
J16570-043  &  1  &  2457822.7113  &2457818.6556  & 1801 &Blue     & V + G & 0.131  &$-109.5 \pm 2.7$  & -165  & $3.4\times10^{16}$ & $2.1\times10^{30}$\\
J18022+642  &  1  &  2458397.4738  &2458396.3735  & 1801 &Red      & V + G & 0.076  & $108.2 \pm 3.2$  &  150  & $9.6\times10^{15}$ & $5.6\times10^{29}$\\
J18482+076  &  1  &  2457631.4686  &2457634.4741  & 1801 &Red      & V + G & 0.192  & $110.7 \pm 5.1$  &  166  & $1.8\times10^{16}$ & $1.1\times10^{30}$\\
J20451-313  &  1  &  2458679.5342  &2458678.5698  & 295 &Red      & L + G & 2.049  &  $23.3 \pm 6.0$  &  353  & $1.6\times10^{19}$ & $4.4\times10^{31}$\\
            &  2  &  2458679.5389  &2458678.5698  & 295 &Blue     & L + G & 3.366  &  $-4.5 \pm 3.0$  & -345  & $2.7\times10^{19}$ & $2.7\times10^{30}$\\
            &  3  &  2459161.2725  &2459154.2966  & 295 &Red      & G + G & 0.727  &  $73.0 \pm 6.8$  &  280  & $5.8\times10^{18}$ & $1.5\times10^{32}$\\
J22114+409  &  1  &  2458599.6441  &2458598.6688  & 1801&Red      & V + G & 4.097  &  $55.5 \pm 7.6$  &  153  & $4.4\times10^{17}$ & $6.7\times10^{30}$\\
J22231-176  &  1  &  2457586.6313  &2457571.6708  & 1800&Red      & V + G & 0.477  & $104.0 \pm 2.2$  &  162  & $7.0\times10^{16}$ & $3.8\times10^{30}$\\
J22468+443  &  1  &  2457632.6332  &2457634.6356  & 119 &Blue     & V + G & 0.133  &$-104.7 \pm 2.7$  & -153  & $7.4\times10^{16}$ & $4.0\times10^{30}$\\
            &  2  &  2457633.4714  &2457634.6356  & 111 &Red      & V + G & 4.242  &  $49.2 \pm 3.6$  &  196  & $2.4\times10^{18}$ & $2.8\times10^{31}$\\
            &  3  &  2457650.5425  &2457647.3787  & 274 &Far-red  & G + 2*G & 1.169 & $168.0 \pm 35.8$ &  330  & $6.5\times10^{17}$ & $9.2\times10^{31}$\\
            &     &                &              & &Red      &         & 0.427 &  $57.5 \pm 8.8$  &  144  & $2.4\times10^{17}$ & $3.9\times10^{30}$\\
            &  4  &  2457968.4267  &2457963.5706  & 545 &Far-red  & G + 2*G & 0.514 & $168.2 \pm 74.9$ &  319  & $2.9\times10^{17}$ & $4.0\times10^{31}$\\
            &     &                &              & &Red      &         & 0.741 &  $73.5 \pm 19.1$ &  169  & $4.1\times10^{17}$ & $1.1\times10^{31}$\\
            &  5  &  2458032.4357  &2458031.6307  & 390 &Red      & V + G & 0.185 & $124.2 \pm 5.0$  &  186  & $1.0\times10^{17}$ & $7.9\times10^{30}$\\
            &  6  &  2458032.6097  &2458031.6307  & 517 &Red      & G + G & 0.679 &  $62.7 \pm 5.5$  &  245  & $3.8\times10^{17}$ & $7.4\times10^{30}$\\
J22518+317  &  1  &  2457762.2794  &2457766.3012  & 705 &Red      & G + G & 1.422 &  $53.8 \pm 2.4$  &  224  & $1.5\times10^{18}$ & $2.1\times10^{31}$\\
J23548+385  &  1  &  2457643.5787  &2457649.5951  & 1800&Blue     & V + G & 0.278 &$-108.9 \pm 3.4$  & -190  & $1.2\times10^{17}$ & $7.4\times10^{30}$\\
\enddata 
\tablecomments{For the fitted functions, "L" represents the Lorentzian profile, "V" denotes the Voigt profile, and "G" signifies the Gaussian profile. For example, "V + G" indicates a Voigt component fitting the symmetric profile combined with a Gaussian component for the asymmetric profile. Additionally, "G + 2*G" means two Gaussian components are used to fit the asymmetric profile.}
\end{deluxetable*}

\section{Discussion}\label{sec4}
\begin{figure*}
\centering
\includegraphics[width=8.5cm,height=6.25cm]{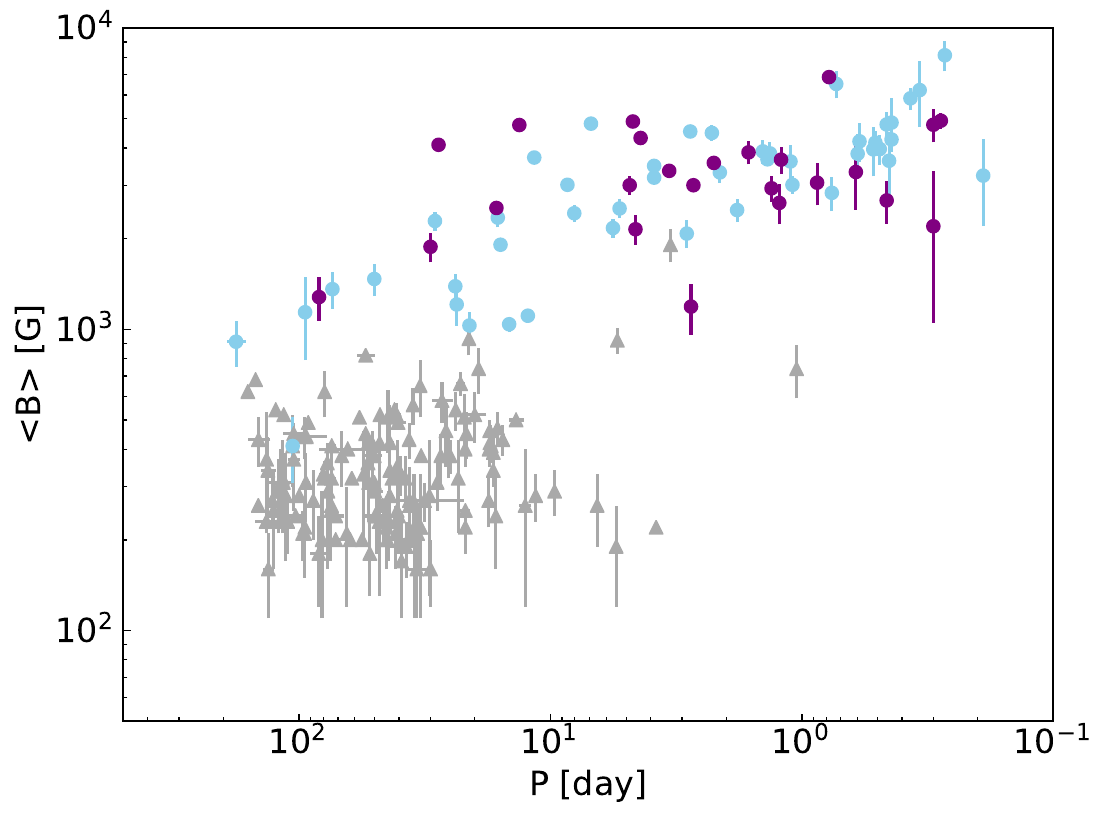}
\includegraphics[width=8.5cm,height=6.25cm]{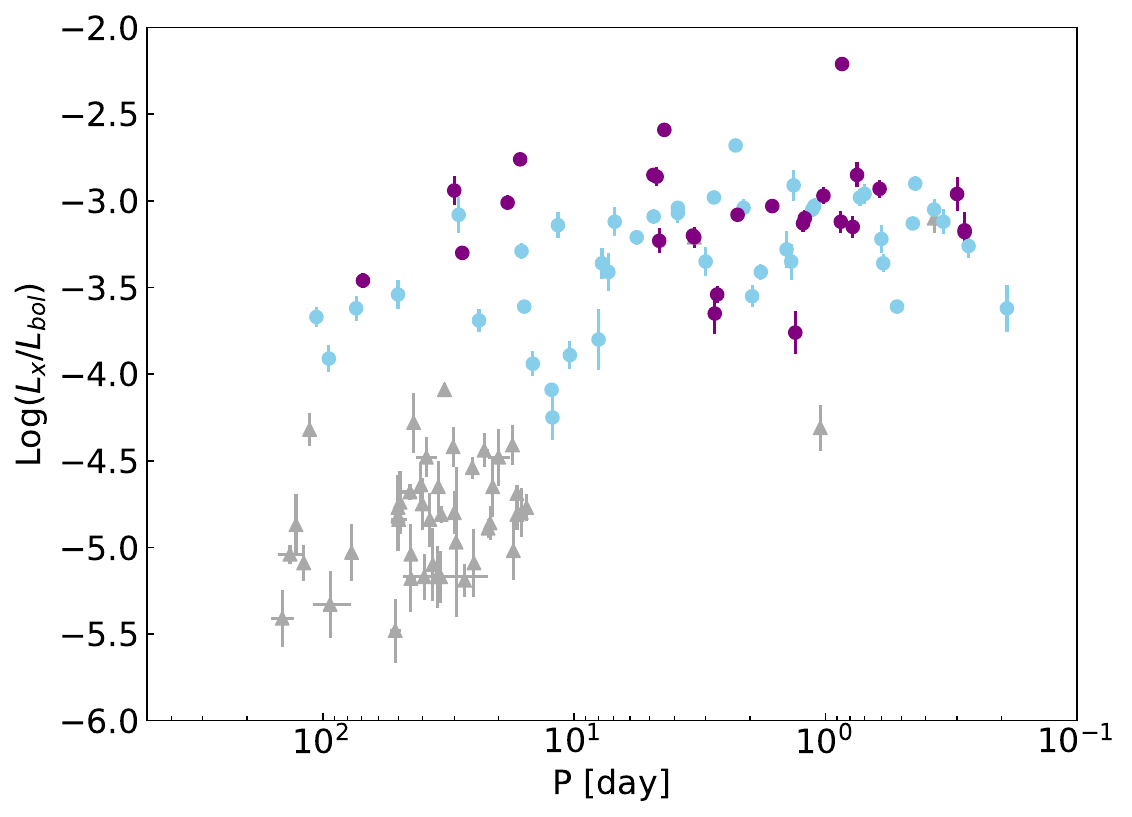}
\includegraphics[width=8.5cm,height=6.25cm]{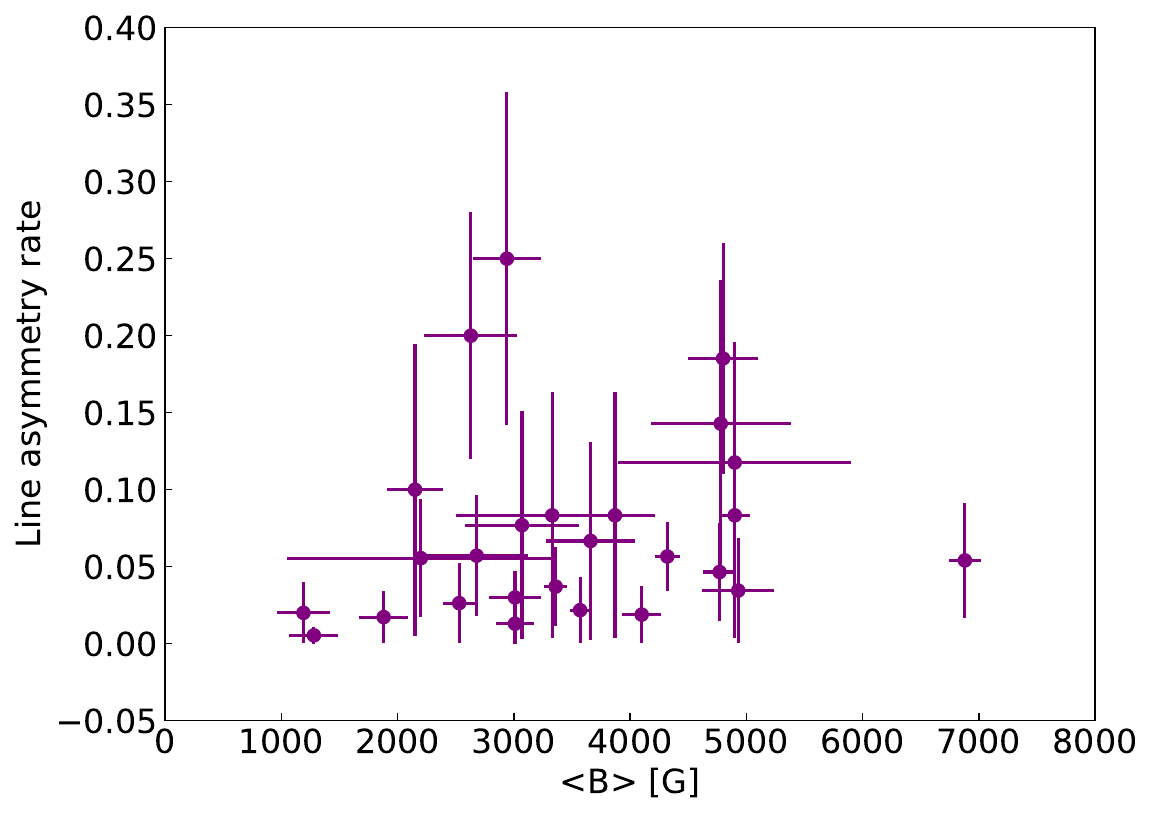}
\includegraphics[width=8.5cm,height=6.25cm]{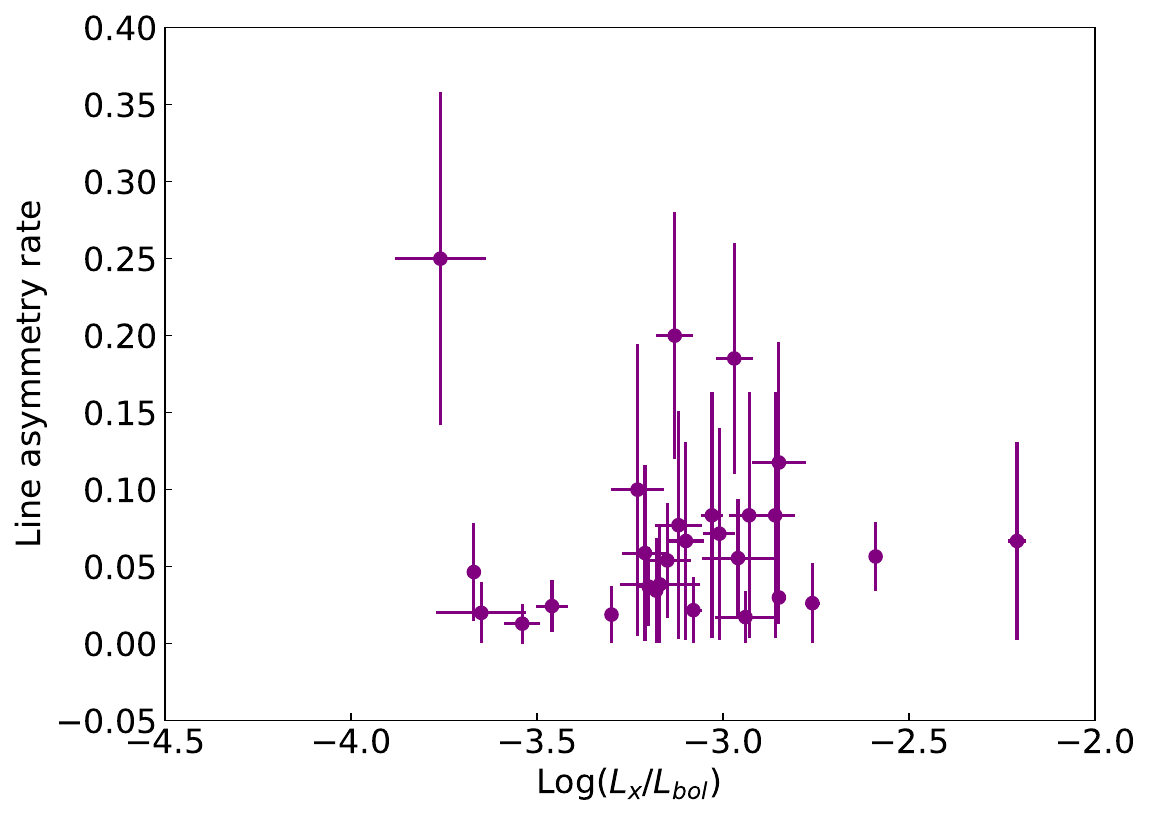}
\caption{Upper panels: The distribution of the surface-average magnetic field ($\langle B \rangle$) (left panel) and normalized X-ray luminosity (log$\frac{L_{X}}{L_{bol}}$) (right panel) vs. the rotational period of the stars under investigation. Gray triangles represent stars that have been excluded from our analysis due to either their inactive nature or low levels of activity. Sky blue dots indicate stars that were examined in our study but do not exhibit asymmetric features in their H${\alpha}$ line profiles. Purple dots denote stars that display asymmetric characteristics in their H${\alpha}$ line profiles. Bottom panels: The distribution of line asymmetry rate vs. $\langle B \rangle$ (left panel) and log$\frac{L_{X}}{L_{bol}}$ (right panel), respectively. Spearman correlation tests show no statistically significant relationships: $\rho = 0.188$ ($p = 0.295$) for $\langle B\rangle$ and $\rho = 0.153$ ($p = 0.397$) for log$\frac{L_{X}}{L_{bol}}$.}
\label{Fig3}
\end{figure*}
\subsection{Potential stellar CMEs}
As described in the previous section, asymmetric features observed in the H${\alpha}$ line profiles are primarily associated with flare-like events. Asymmetries  can effectively trace moving plasma during stellar flares, particularly in serving as crucial indicators of prominence and filament eruptions, as well as potential stellar CMEs. Blueshifted emission asymmetries are typically linked to prominences or filaments \citep[the latter as indicated by][]{Leitzinger2022MNRAS} that are erupting towards the observer. In contrast, redshifted emission asymmetries indicate backward-directed prominence eruptions occurring near the stellar limb. Consequently, the asymmetric features we have detected in the M-type dwarfs can be attributed to prominence or filaments eruptions.

In our analysis, we find that the bulk velocities of these asymmetric features range from tens of km~s$^{-1}$ to 250 km~s$^{-1}$, while their maximum velocities are with the order of 100--450 km~s$^{-1}$. Both the bulk and maximum velocities are smaller than the escape velocities ($\sim$~400--650 km~s$^{-1}$, see Table~\ref{tab1}) for our investigated stars. An exception is noted for the star RBS~365 (Karmn~J02519+224), its maximum velocity of far-blueshifted asymmetric profile ($\sim$~404 km~s$^{-1}$) is comparable to this star's escape velocity ($\sim$~403 km~s$^{-1}$). The results of low velocities align with previous study conducted by \citet{Vida2019}, who reported asymmetries characterized by typical observed maximum velocities in the range of 100 to 300 km~s$^{-1}$. Notably, several stars have been analyzed in both our study and theirs, particularly V388~Cas (Karmn~J01033+623) and EV~Lac (Karmn~J22468+443), which displayed significant asymmetric features. 

It is important to note that the velocity measurements represent only components along the line of sight due to projection effects. This is particularly significant for prominence eruptions occurring near the stellar limb. Consequently, these prominence eruptions could exhibit significantly higher true eruption velocities and have the potential to evolve into stellar CMEs. Moreover, as discussed by \citet{Vida2019}, there are additional potential reasons for the situations. For instance, we may only be able to observe early phases of CMEs that display lower velocities. Furthermore, strong magnetic fields might inhibit prominence or filament materials from escaping off the stellar surface \citep{Drake2016, Alvarado2018}. The combination of these scenarios may collectively contribute to the observed lower velocities of asymmetric features.

The masses of these potential stellar CMEs fall within the range of $10^{15}$ to $10^{19}$~g, while their kinetic energies are approximately in the range of $10^{29}$ to $10^{32}$~erg. The mass estimates are consistent with the findings reported by \citet{Vida2019} and \citet{Koller2021}, who determined that the typical masses of the ejecta for M-type dwarfs were in the order of \(10^{15}\)--\(10^{18}\) g and \(10^{16}\)--\(10^{18}\) g, respectively. Notably, these masses basically exceed that of solar CMEs, which have an average mass around $10^{14}$~g and a maximum mass near $10^{17}$~g \citep{Gopalswamy2009}. In contrast, the kinetic energy values are basically comparable to those observed in solar CMEs, which exhibit an average kinetic energy of about $10^{29}$~erg and a peak value close to $10^{33}$~erg \citep{Gopalswamy2009}. However, it is crucial to emphasize that the kinetic energies calculated for these potential stellar CMEs represent lower bounds. This limitation arises from the projection effects of velocities and the constraints on mass estimates.

However, stellar CMEs are not the only cause of line profile asymmetries; other plasma motions during flares can also contribute to them. Current knowledge taken from solar observations indicates that blueshifted emission asymmetries may result from chromospheric evaporation \citep[e.g.][]{Tei2018}, typically reaching velocities of several tens of km~s$^{-1}$. In contrast, redshifted emission asymmetries might suggest chromospheric condensation, coronal rain along post-flare loops, or falling prominence materials \citep{Koller2021, Wu2022}. Coronal rain descends at speeds between 30 and 200 km~s$^{-1}$, averaging around 60 to 70 km~s$^{-1}$ \citep{Antolin2012, Lacatus2017}. Chromospheric condensation is generally observed with speeds of several tens of km~s$^{-1}$ \citep{Ichimoto1984}. In stellar case, \citet{Wollmann2023} conducted a modeling study of coronal rain along H${\alpha}$-emitting post-flare loops for the M-type dwarf AD Leo. Their findings demonstrate a strong agreement with observations of redshifted asymmetries.

\subsection{Statistical Properties of M-type dwarfs with CME candidates and line asymmetries}
\citet{Reiners2022} provided direct measurements of the average magnetic field strengths for 292 M-type dwarf stars within the CARMENES survey. In conjunction with values obtained from \citet{Shulyak2019}, a significant proportion of the stars we examined has such average magnetic field measurements, thereby enabling a comprehensive statistical analysis of their magnetic properties. As illustrated in Figure~\ref{Fig3}, stars with lower  surface-average magnetic field ($\langle B \rangle$) are largely excluded from our analysis. This exclusion is based on a visual inspection of their H${\alpha}$ line characteristics (see Section~2 for details). Notably, among the stars investigated, asymmetric features in their spectral profiles can be detected irrespective of whether they exhibit strong or relatively weak $\langle B \rangle$. However, most features are primarily concentrated in stars with higher $\langle B \rangle$, which also tend to be the stars rotating faster than the saturation period of approximately 10~days \citep{Reiners2022}. Moreover, the line asymmetry rate—defined as the ratio of spectra exhibiting asymmetrical features to the total number of analyzed spectra for each star, with the error bars representing the standard error of this proportion, estimated as $\sqrt{p(1-p)/N}$ (where $p$ is the asymmetry rate and $N$ is the total number of spectra analyzed for the star)—shows no statistically significant correlation with the $\langle B \rangle$ (Spearman's $\rho = 0.188$, $p = 0.295$). This lack of correlation may be due to the saturation effect in these stars. Additionally, \citet{Alvarado2018} considered a set of three-dimensional MHD numerical simulations to study the suppression of CMEs by large-scale magnetic field and found that a large-scale dipolar magnetic field with a strength of 75~G is capable of fully confining eruptions within the stellar corona. It is noteworthy that the $\langle B \rangle$ of our investigated stars is significantly stronger than this specific value (see Table~\ref{tab1}). Consequently, this might raises an important question: Can these ejections truly escape from the stars with such strong magnetic field? Further research is necessary to address this inquiry.

In addition, Figure~\ref{Fig3} illustrates the distribution of the normalized X-ray luminosity, log$\frac{L_{X}}{L_{bol}}$, for the stars included in our study. Given that stellar CMEs originate from the corona, it is a well-established assumption that CMEs are closely linked to coronal activity. Our analysis reveals that stars excluded based on their H${\alpha}$ characteristics also exhibit lower log$\frac{L_{X}}{L_{bol}}$. Notably, stars displaying asymmetric features in their H${\alpha}$ profiles tend to show higher log$\frac{L_{X}}{L_{bol}}$ and are situated within the saturation regime of stellar activity, which corresponds to the log$\frac{L_{X}}{L_{bol}}$ range of approximately -3.5 to -3.0 \citep[e.g.,][]{Mathioudakis1995, Stauffer1997ApJ}. Moreover, primarily due to the saturation of X-ray activity, the line asymmetry rate also shows no significant correlation with log$\frac{L_{X}}{L_{bol}}$ (Spearman's $\rho = 0.153$, $p = 0.397$). \citet{Vida2019} investigated the relationship between line asymmetry rate and various physical parameters of the stars under investigation during their search for stellar CMEs. Their findings indicate that line asymmetry rate increases after surpassing a threshold of X-ray activity, which coincides with a saturation of activity levels. Additionally, they noted that the correlation between the asymmetry rate and the absolute X-ray luminosity log$L_{X}$ was weaker than those with other activity indicators. Our results are largely consistent with theirs. However, because we preemptively excluded inactive and low-activity stars from our sample, no such threshold is evident in our findings.

\section{Summary and Future works}\label{sec5}
In this study, we employ the CARMENES DR1 spectra to investigate potential stellar CMEs in M-type dwarfs by analyzing asymmetries of the H${\alpha}$ profiles during flare events. We have identified 61 distinct asymmetric features on 33 M-type dwarfs. These asymmetric features are predominantly associated with enhanced H${\alpha}$ emissions, as well as increased emissions in the $\mbox{He~{\sc i}}$ D$_{3}$, $\mbox{Ca~{\sc ii}}$ $\lambda$8498, $\mbox{Ca~{\sc ii}}$ $\lambda$8542, and $\mbox{Ca~{\sc ii}}$ $\lambda$8662 lines, suggesting probable connections to flare events. The inferred velocities of these asymmetries remain below the respective stellar escape velocities, with the exception of one case in which the maximum velocity approaches the escape velocity. While these features could be indicative of potential stellar CMEs, alternative plasma motions during flares must also be considered. The estimated masses of the moving plasma responsible for these asymmetries range from $10^{15}$ to $10^{19}$~g, with corresponding kinetic energies of $10^{29}$--$10^{32}$~erg. The velocity and mass ranges are basically consistent with the findings derived by previous studies for M-type dwarfs. For most of our sample, measurements of the surface-average magnetic field ($\langle B \rangle$) and the normalized X-ray luminosity (log$\frac{L_{X}}{L_{bol}}$) are available. We find that stars exhibiting asymmetric features in their H${\alpha}$ profiles tend to manifest stronger average magnetic fields and higher normalized X-ray luminosity. However, no significant correlation is observed between the line asymmetry rate and both the $\langle B \rangle$ and log$\frac{L_{X}}{L_{bol}}$, as quantified by Spearman correlation tests.
 
In the future works, we intend to provide a detailed investigation on specific asymmetric features observed in our target stars—particularly those exhibiting complex profiles and those having simultaneous observations from the TESS. Furthermore, due to challenges in ruling out alternative processes that could account for the observed asymmetries and difficulties in determining whether the ejected materials ultimately escape from host stars, it is currently impossible to confirm that these asymmetrical features can be definitively attributed to CMEs. Therefore, our future studies will involve time-series spectroscopic observations of several intriguing target stars, aimed at investigating both intensity and velocity evolution associated with potential asymmetrical features. This will significantly enhance our ability to identify stellar CMEs more effectively.

\begin{acknowledgments}
This work is supported by the National Natural Science Foundation of China under grant No. 12288102. We also thank the financial support from the National Natural Science Foundation of China under grant Nos. 10373023, 10773027, U1531121, 11603068 and 12373039, the China Manned Space Program with grant No. CMS-CSST-2025-A15, the Yunnan Revitalization Talent Support Program (Young Talent Project), International Centre of Supernovae, Yunnan Key Laboratory (grant No. 202302AN360001), and the Yunnan Fundamental Research Project (grant No. 202305AS350009).
\end{acknowledgments}

\bibliography{sample701}{}
\bibliographystyle{aasjournalv7}



\end{document}